\def\paperversion{camera}
\def\grammarly{off}
\pdfoutput=1

\def\paperversiondraft{draft}
\def\paperversionblind{blind}
\def\paperversioncamera{camera}
\ifx\paperversion\paperversiondraft
\else
  \ifx\paperversion\paperversionblind
  \else
    \ifx\paperversion\paperversioncamera
    \else
       \errmessage{paperversion must be set to 'draft', 'blind', or 'camera'}
    \fi
  \fi
\fi

\def\grammarlyon{on}

\ifx\grammarly\grammarlyon
\def\review{}
\else
\def\review{review,}
\fi

\ifx\paperversion\paperversiondraft
  \documentclass[nonacm,sigplan,\review,screen]{acmart}
\fi

\ifx\paperversion\paperversionblind
  \documentclass[nonacm,sigplan,\review anonymous,screen]{acmart}
\fi

\ifx\paperversion\paperversioncamera
  \documentclass[nonacm,sigplan,screen]{acmart}
\fi

\usepackage{xargs}
\usepackage{todonotes}
\usepackage{xparse}
\usepackage{xifthen, xstring}
\usepackage[normalem]{ulem}
\usepackage{pifont}
\usepackage{xspace}
\usepackage{subfig}
\usepackage{listings}
\usepackage{enumerate}
\usepackage[inline]{enumitem}
\setlist{nosep}
\setlist{noitemsep}

\makeatletter
\font\uwavefont=lasyb10 scaled 652
\newcommand\colorwave[1][blue]{\bgroup\markoverwith{\lower3\p@\hbox{\uwavefont\textcolor{#1}{\char58}}}\ULon}
\makeatother

\ifx\paperversion\paperversiondraft
\newcommand\createtodoauthor[2]{%
\def\tmpdefault{emptystring}
\expandafter\newcommand\csname #1\endcsname[2][\tmpdefault]{\def\tmp{##1}\ifthenelse{\equal{\tmp}{\tmpdefault}}
   {\todo[linecolor=#2!20,backgroundcolor=#2!25,bordercolor=#2]{\textbf{#1:} ##2}}
   {\ifthenelse{\equal{##2}{}}{\colorwave[#2]{##1}\xspace}{ \todo[linecolor=#2!10,backgroundcolor=#2!25,bordercolor=#2]{\textbf{#1:} ##2}\colorwave[#2]{##1}}}}}
\else
\newcommand\createtodoauthor[2]{%
\expandafter\newcommand\csname #1\endcsname[2][\@nil]{}}
\fi

\ifx\paperversion\paperversiondraft
  \paperwidth=\dimexpr \paperwidth + 6cm\relax
  \oddsidemargin=\dimexpr\oddsidemargin + 3cm\relax
  \evensidemargin=\dimexpr\evensidemargin + 3cm\relax
  \marginparwidth=\dimexpr \marginparwidth + 3cm\relax
\fi

\createtodoauthor{KK}{red}
\createtodoauthor{IBM}{blue}
\createtodoauthor{TG}{green}
\createtodoauthor{TODO}{orange} 
\createtodoauthor{AS}{purple}
\createtodoauthor{NI}{yellow}

\makeatletter\if@ACM@journal\makeatother
\acmJournal{PACMPL}
\acmVolume{1}
\acmNumber{1}
\acmArticle{1}
\acmYear{2017}
\acmMonth{1}
\acmDOI{10.1145/nnnnnnn.nnnnnnn}
\startPage{1}
\else\makeatother
\acmConference[PL'17]{ACM SIGPLAN Conference on Programming Languages}{January 01--03, 2017}{New York, NY, USA}
\acmYear{2017}
\acmISBN{978-x-xxxx-xxxx-x/YY/MM}
\acmDOI{10.1145/nnnnnnn.nnnnnnn}
\startPage{1}
\fi

\setcopyright{none}             

\newcommand{\Sec}[1]{\S{\ref{#1}}}
\newcommand{\Fig}[1]{Fig.~\ref{#1}}

\newcommand{\Lst}[1]{Listing~\ref{#1}}

\begin{document}


\title{Compiling Neural Networks for \\
    a Computational Memory Accelerator}


\author{Kornilios Kourtis}
\authornote{Majority of work done while at IBM Research}
\affiliation{
  \position{Independent researcher}
}

\author{Martino Dazzi}
\affiliation{
  \position{IBM Research}
}

\author{Nikolas Ioannou}
\affiliation{
  \position{IBM Research}
}

\author{Tobias Grosser}
\affiliation{
  \position{ETH Zurich}
}

\author{Abu Sebastian}
\affiliation{
  \position{IBM Research}
}

\author{Evangelos Eleftheriou}
\affiliation{
  \position{IBM Research}
}

\begin{abstract}
Computational memory (CM) is a promising approach for accelerating inference on
neural networks (NN) by using enhanced memories that, in addition to storing data,
allow computations on them.
One of the main challenges of this approach is defining a hardware/software
interface that allows a compiler to map NN models for efficient execution on
the underlying CM accelerator. This is a non-trivial task because efficiency
dictates that the CM accelerator is explicitly programmed as a
dataflow engine where the execution of the different NN layers form a
pipeline.

In this paper, we present our work towards a software stack for executing ML
models on such a multi-core CM accelerator.  We describe an architecture for
the hardware and software, and focus on the problem of implementing the appropriate
control logic so that data dependencies are respected. We propose a solution to
the latter that is based on polyhedral compilation.
\end{abstract}

\maketitle
\ifx\grammarly\grammarlyon
\onecolumn
\else
\fi

\section{Introduction}

As  general-purpose architectures hit scalability limits, important applications
such as machine learning (ML) turn to specialized hardware to meet their energy
and performance requirements~\cite{jouppi2017datacenter, sze2017efficient,
moreau2018hardwaresoftware, chen2016eyeriss, kwon2018maeri, nvdla,
chen2014diannao, parashar2017scnn, armml-hotchips18}.

Computational memory (CM) contrasts traditional Von Neumann
architectures, which separate computation and memory, by enabling memory
to perform computations on the data it holds. Specifically, technologies such as
PCM or Flash can be used to build crossbar arrays that, using Kirchhoff’s laws,
implement an analog Matrix-Vector multiplication (MxV), where the matrix data are
stored in the crossbar memory
cells~\cite{eleftheriou2019inmem,mythic-hotchips18,ankit2019puma}.
Such a crossbar can execute an MxV in a single step, whereas digital logic
typically requires multiple steps. At the same time, combining compute and
storage in a single unit reduces communication which constitutes the main
challenge of data-intensive workloads. This makes CM an attractive alternative
for accelerating ML workloads~\cite{song2017pipelayer, liu2015reno,
jain2018rxnn, cheng2017time, yu2018neuro, li2017drisa,
fujiki2018memory,shafiee2016isaac, ankit2019puma, ambrosi2018hardware,chi2016prime}.

CM technologies such as PCM or Flash require significant time to program
(write).
This renders the common practice of using accelerators to execute one NN layer
at a time~\cite{nvdla, chen2016eyeriss, chen2014diannao, jouppi2017datacenter,
chen2014dadiannao} impractical for these technologies due to the high overhead
of reconfiguring the crossbars, which might take as long as a
minute~\cite{mythic-hotchips18}.
Instead, efficiency dictates that the CM accelerator is configured once to
implement a given NN. After configuration, inference is performed by streaming
input data to the accelerator.

In this work, we discuss our approach towards a software stack that targets such
a multi-core CM chip for accelerating deep learning inference at the
edge~\cite{eleftheriou2019inmem,Y2019sebastianJPD,Y2019sebastianVLSI}, where
each core includes a crossbar that implements an analog MxV operation. The
accelerator follows a dataflow processing model, where inference is executed as
a pipeline formed by the different NN layers.

Our ultimate goal is to build a software stack that enables transparent use of
the CM accelerator, and, at the same time, guide hardware design so the
accelerator can be better utilized by software. Hence, we design the compiler
and the rest of the software stack in tandem with the accelerator.
Specifically, we prototype a Computational Memory Neural Network Compiler
(cmnnc) that aims to compile NN models to be executed on the CM accelerator, and
a simulator that models such hardware and acts as the target platform.

A key problem we faced, that stems from the fact that the NN network needs to be
fitted into the accelerator, is generating the
control logic between the CM cores that execute different layers of the NN so
that data dependencies are respected. Because traditional accelerators do not
require this feature, existing ML compilers~\cite{rotem2018glow, chen2018tvm,
leary2017xla} do not have facilities for tackling it.
We address this challenge by using polyhedral compilation techniques
\cite{feautrier1992somea, feautrier1992someb} to represent data dependencies and
generate code for state machines that implement the desired control.

In summary, the \emph{contributions} of this work are:
\begin{enumerate*}[label={\bfseries\arabic*)}]
  \item We propose an architecture for an inference CM accelerator and the software stack
   for driving it.

  \item We discuss an approach for compiling NN models for the proposed CM
  accelerator, focusing on using polyhedral compilation to express the control
  logic between the CM cores so that data dependencies are respected. To the
  best of our knowledge this is an open problem, and we are the first to use
  polyhedral compilation to tackle it.
\end{enumerate*}

While we believe that our approach is generally applicable to dataflow
architectures, for the sake of brevity and clarity our discussion assumes a
specific architecture which is described in \Sec{sec:hw}.  In \Sec{sec:cmnnc}
we present the proposed compiler, and a solution for dealing with data
dependencies. Finally, we discuss related work in \Sec{sec:related}, and
conclude in \Sec{sec:conclusion}.

\section{The Computational Memory Accelerator}
\label{sec:hw}

We start by  discussing a functional model of the hardware. This model is meant
as a vehicle for co-designing the hardware interface with the software
stack, and hence omits many details about the hardware implementation.

\begin{figure}[tb]
    \centering
    \includegraphics[width=\columnwidth]{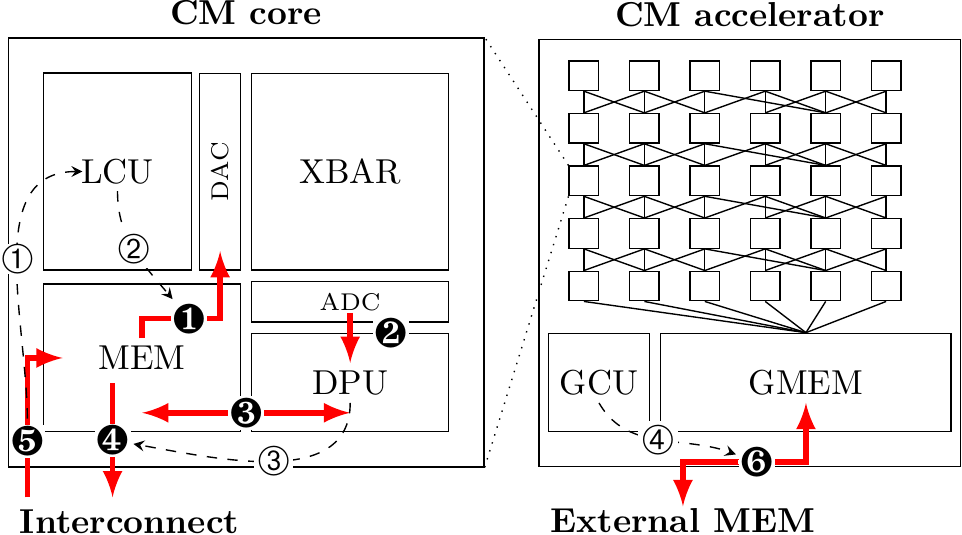}
    \caption{CM core and CM accelerator architecture.}
    \label{fig:arch}
\end{figure}

The left side of \Fig{fig:arch} shows the various components of the CM core.
In addition to the crossbar (XBAR), the core also includes a lightweight
digital processing unit (DPU), and local memory (MEM), which is, typically, a
few kilobytes of SRAM.
The crossbar array implements an analog MxV operation, where the matrix M data
(typically, weights) are stored directly in the crossbar's memory cells, while
vector V (typically, activations) is loaded from the local memory.
It is worth noting that the crossbar has certain dimensions reflected in the
size of the MxV operations. We call this dimension the \emph{width} of the
unit, i.e., a width of 64 means that $M$ contains $64\times64$ elements, while
input and output vectors contain 64 elements.

The motivation behind this
design is to run inference on NNs by executing each different layer on a
separate core, thus forming a pipeline between the cores that resembles the
structure of the NN. The operation that we target to accelerate (e.g., the
convolution operation on convolutional NNs) executes on the crossbar, while
everything else (e.g., activation functions, pooling layers) executes on the DPU.

In \Fig{fig:arch}, thick red lines and labels with black background denote data
operations while dashed lines and labels with white background denote control
operations.  Execution proceeds in cycles.  During a cycle, the local control
unit ({LCU}) loads the data of the input vector from the local SRAM to the
crossbar (\ding{193}, \ding{182}) so that the analog MxV is performed. When the
MxV operation completes, its output vector is made available to the DPU
(\ding{183}), which then executes a sequence of instructions. During the execution,
the DPU may load and store data to the local SRAM (\ding{184}), and schedule
data transfers from the local SRAM to other cores (\ding{195}, \ding{185}). The
data will become available on the remote core's local SRAM on the next cycle
(\ding{186}).

The cores are organized in a pipeline, where the input of one is the output of
another. Hence, in many cases one core has to wait until all necessary data
from remote cores become available before executing. To this end, the LCU
``snoops'' the remote writes to SRAM (\ding{192}, \ding{186}) and implements a
state machine that controls execution. Depending on this state machine, the
LCU may either do nothing or execute a computation step. In the latter
case, specific SRAM values are loaded to the crossbar input (\ding{193}), the
MxV operation is executed, and then DPU executes its instructions.

The CM accelerator chip (right side of \Fig{fig:arch}) includes a number of interconnected
cores, a global input/output buffer (GMEM), and a global control unit (GCU).
The GCU issues DMA operations for transferring data between the external (e.g.,
host) memory and the chip's GMEM (\ding{195}, \ding{187}).  The GCU also
transfers data from (to) the GMEM to (from) cores that act as input (output)
nodes in the dataflow graph.

The hardware model of the CM accelerator also includes the topology of the
interconnect. A simple approach would be to assume that all cores are connected
with each other, either via an all-to-all topology or by having a routing
scheme that enables all-to-all communication via message forwarding. Both of
these approaches, however, have inefficiencies~\cite{kwon2017rethinking}, and
we, instead, decide to expose the interconnect topology to the compiler so that
it can optimize the mapping of NN layers to CM cores accordingly.
We represent the chip topology as a directed graph, where an edge from one unit
to another means that the first can send data to the second.  An example of an
effective interconnect topology for a CM accelerator is described by Dazzi et
al.~\cite{dazzi20195}.

\section{Compiling NNs for the CM accelerator}
\label{sec:cmnnc}

Machine learning frameworks such as Tensorflow or PyTorch represent NNs
as dataflow graphs where nodes are computational operators (e.g., a
matrix multiplication), and edges are data dependencies between the operators.
This approach is well-suited for mapping these networks onto the CM accelerator
that also follows a dataflow execution model.
Inference is performed on pre-trained models that bundle two types of data: the
dataflow graph, and initialization data for the operators (e.g., model weights)
as produced by a separate training phase. 
Abstractly, the role of the compiler is to map such a model (in our prototype we
use the ONNX format~\cite{bai2019}) to the CM accelerator for execution.

We focus on convolutional NNs (CNNs), NNs based on convolutions,
which is where most of their execution
time is spent~\cite{cong2014minimizing}. We accelerate these networks by mapping the
convolution to the crossbar's MxV operation (see~\Lst{lst:convmxv}), and use
the DPU for the remaining operations.

\begin{lstlisting}[float,
                 caption={Convolution using an MXV operation (implemented via Python NumPy).},
                 label=lst:convmxv]
# input (inp) shape: (D,IH,IW)
# filters (flt) shape: (FL,D,FH,FW)
# output (out) shape: (D,OH,OW)
def conv2d_mxv(inp,flt,out):
  N = FD*FH*FW
  m = flt.reshape((FL,N))
  for oh in range(OH):
    for ow in range(OW):
      v = inp[:,oh:oh+FH,ow:ow+FW].reshape(N)
      out[:,oh,ow] = matmul(m,v)
\end{lstlisting}

From a software perspective, there are three phases in using the CM
accelerator: compilation, initialization, and execution. Compilation uses the
NN dataflow graph and a hardware description of the accelerator
(number of cores and their properties, interconnect topology, etc.) as input, and
produces a configuration for each individual unit of the accelerator (GCU, DPUs,
LCUs).  These configurations, bundled together and serialized,
initialize the accelerator. After initialization, inference on the compiled
model is executed by streaming data to the accelerator.

Compilation is performed in two steps: \emph{partitioning}, where the dataflow
graph is partitioned so that each partition is mapped to a different CM core,
and \emph{lowering}, where the compiler processes each partition separately and
produces the configuration for each of the units of the corresponding core.
Note that the partitioning step must adhere to the constrains imposed by the
hardware. For example, local objects have to fit into local memory, and if two
dataflow nodes connected via an edge are mapped to different cores, the cores
should also be connected in the hardware interconnect graph.

After compilation, the serialized configurations are used together with the
model weights to initialize the CM accelerator. Model
weights are used for programming the crossbar arrays, but also to initialize
objects that reside in accelerator-local memories.  Once the accelerator is
initialized, the execution phase may start where an external process (e.g., a
driver running on the host) passes descriptors for the input and output data to
the GCU.

Next, we discuss our approach for building such a compiler. Our ideas are
realized in a prototype implementation called cmnnc (Computational
Memory Neural Network Compiler).

\subsection{Partitioning and Mapping}

\begin{figure}[t!]
    \centering
    \includegraphics[width=0.8\linewidth]{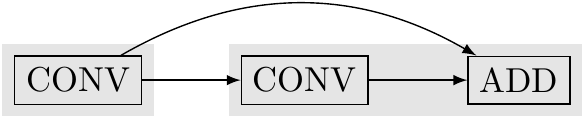}
    \caption{Dataflow graph example with two convolutions and an addition node.}%
    \label{fig:convresidual}
\end{figure}

As discussed previously, the dataflow
graph of the NN is partitioned during the first phase of the compilation, and each partition is mapped onto a different CM core on
the accelerator. We perform these two steps, namely partitioning and mapping, separately. For the first step, we
enforce two invariants: that each partition has \emph{at most} one convolution operator (or
more generally speaking, one operator executed using the crossbar), and that there are no
cycles in the partition graph. These invariants are not necessary, but they
simplify the compilation problem without limiting applicability in practice.
Consider, for example, the dataflow graph of \Fig{fig:convresidual}, with two
convolution and one addition operators. Following the two invariants
presented above the graph is partitioned as shown in the figure: based on the
first invariant we must create two partitions, one for each convolution
operation.  Based on the second invariant, we must bundle the addition
operation with the right hand-side partition; if we bundle it with the left
hand-side partition, a cycle is formed and the invariant breaks.

We perform the partitioning by iterating the dataflow nodes in their topological
order, and creating a new partition whenever we encounter
a convolution node. (This assumes that there are no cycles in the dataflow
graph, but this is a common assumption, e.g., the ONNX format also disallows
cycles.) Given a set of partitions, edges of the dataflow graph are either
within the same partition or span multiple partitions. The latter type of edges
define the partition graph. We map the partition graph to the CM accelerator,
i.e., mapping each partition to a CM core and each edge to a connection in the
interconnect topology, by expressing the problem as a set of constrains in the
Z3 SMT solver~\cite{de2008z3}.

\subsection{Lowering}

Once the partitions are defined and mapped to CM cores, the compiler produces
the configurations for the GCU, LCUs, and DPUs (lowering phase).

The problem of configuring the DPU, i.e., defining the set of instructions to
execute after the MxV operation, is very similar to the problem solved by
existing ML compiler
frameworks~\cite{rotem2018glow,chen2018tvm,leary2017xla},
and in practice we expect to address it by developing an appropriate backend
for such a framework that targets the DPU instruction set.

Configuring the GCU and LCUs, however, cannot, to the best of our knowledge, be
addressed by existing ML compiler frameworks, even the ones that target
specialized ML accelerator hardware. The reason is that existing ML
accelerators typically provide offloading of certain operations, but are not
explicitly programmed as a dataflow engine, so such functionality is not
needed.
As we discussed previously, however, explicitly mapping the NN execution into a
dataflow graph on the CM accelerator is necessary due to the high cost of
reconfiguring the crossbars.
In the next paragraphs, we describe the problem of configuring the LCU, and
propose a solution that utilizes polyhedral compilation. A similar approach can
be used for the GCU.

As is the case with existing frameworks, we represent the inputs and outputs of
operations as tensors, i.e., multi-dimensional arrays. These arrays reside on
the local memory of the CM core that reads them. In the context of the LCU, we
are concerned with arrays that are written to and read by different cores, i.e.,
the arrays that are defined by cross-partition edges in the partition dataflow
graph. For example, in~\Fig{fig:convresidual}, the output of the first
{\ttfamily CONV} operator on the first partition is read by the operators of the
second partition.

The LCU executes a state machine that snoops the remote writes from other cores
to the local objects and decides when and how (e.g., what data to load to the
crossbar) to trigger the local computation. For example, if we consider
\Lst{lst:convmxv}, the first iteration can be performed only when the data in
{\ttfamily inp[:,0:FH,O:FW]} have been written (either by the GCU or by
another core). In other words, there is a read after write (RAW) dependency that
needs to be respected to ensure correct execution. Next, we discuss how we can
generate the LCU state machine using the polyhedral model.

\subsection{Enforcing dependencies via the polyhedral model}
\label{sec:deps}

To formally reason about data dependencies across operations and partitions, we use the polyhedral model, where
computations are represented as nested loops that access multi-dimensional
arrays. Since most nodes on NN dataflows are linear algebra  operators with
tensor operands, the polyhedral model works well for such applications.
Specifically, we use the integer set library
(ISL)~\cite{verdoolaege2016presburger,verdoolaege2010isl} to represent the
properties of computations as Presburger sets and relations.  ISL allows
representing sets of integer tuples, and relations that map elements
of one set to another, efficiently, without enumerating all their elements.  ISL sets are used
to model iteration spaces, where each tuple consists of the values of the induction
variables of a loop nest, as well as array locations, where each tuple is a
multi-dimensional index into the array elements. ISL relations are used to
expressed mappings between such sets (e.g., data dependencies).

\begin{figure}[t!]
    \centering
    \includegraphics[width=.99\linewidth]{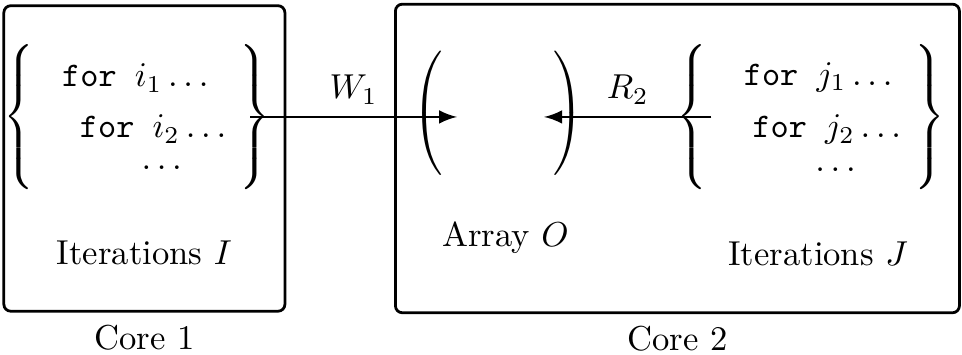}
    \caption{Modeling dependencies between CM cores.}%
    \label{fig:deps}
\end{figure}

To illustrate these concepts, we consider two cores, each executing its own loop
nest and an array $O$ that the first core writes to and that the second reads from
(\Fig{fig:deps}). (For simplicity, we ignore the outermost loop, which is
iterating over input data and is generally unbounded.)
Each core's loop nest defines an instance set: $(i_1, i_2, \ldots) \in I$ for
core 1, and $(j_1, j_2, \ldots) \in J$ for core 2.
The data each iteration reads (writes) from (to) arrays defines the read (write)
\emph{access relation}, which maps each iteration instance to one or more array
locations.
For example, if $R_2$ is the read relation of $J$ to $O$ ($J \rightarrow O$),
$R_2$ includes a $(j \rightarrow o)$ pair \emph{iff} iteration $j \in J$
reads from location $o \in O$.
\Lst{lst:rdacc} shows the read access relation for array {\ttfamily inp} of
\Lst{lst:convmxv}, where tuples of values for the induction variables
({\ttfamily CONV\_MXV[oh,ow]}) are mapped to locations on the input array
({\ttfamily inp[id,ih,iw]}) using inequality constrains.
Similarly, if $W_1$ is the write relation of $I$ to $O$ ($I \rightarrow O$), then
$W_1$ includes a $(i \rightarrow o)$ pair \emph{iff} iteration $i \in I$
writes to location $o \in O$.
We assume that object locations are written to at most once, i.e., write relations
are injective. Note that we do not (indeed cannot) make the same
assumption for read relations.

\begin{lstlisting}[
    language={},%
    label=lst:rdacc,
    float=t!,%
    caption={Read access relation for \Lst{lst:convmxv}}%
]
{ CONV_MXV[oh,ow] -> inp[id,ih,iw] :
      0 <= oh < OH
  and 0 <= ow < OW
  and 0 <= id < D
  and oh <= ih < oh + FH
  and ow <= iw < ow + FW }
\end{lstlisting}

We can enforce dependencies by executing all iterations of $I$ on core 1 before all
iterations of $J$ on core 2, but this is inefficient. Instead, we
want to allow cores to execute iterations in parallel as much as possible, forming a pipeline.
Hence, our goal is to compile a state machine that observes writes from
iterations in $I$, and advances iterations in $J$ so that they are only executed
if all the data they read have already been written.
To this end, we use the ISL algebra to compute relation $\mathcal{S} \enspace
(O \rightarrow J)$ that maps observed writes in $O$, to the maximum (based on
execution order) iteration in $J$ that can be executed. We present the  steps
for computing $\mathcal{S}$ in ISL in Appendix~\ref{app:isl}.

Using relation $\mathcal{S}$, we can generate code for the LCU state machines.
Each cross-partition edge in the dataflow graph defines an array shared by two
partitions: its writer (source) and its reader (destination). We compute the
access relations (read and write) for each array based on the operator type
(e.g., convolution) and parameters (e.g., convolution kernel size, padding,
etc.). Note that we can combine edges with the same source and destination, so
that only one array is used. For example, for the graph of \Fig{fig:convresidual}
we use a single array by combining the read access relations of the {\ttfamily
CONV} and {\ttfamily ADD} node. For every partition, we compute the
relation $\mathcal{S}$ for every object read. Using these relations, we use the
ISL AST facilities to generate code that implements the LCU state
machine.

\subsection{Prototype}

We realize our approach in a prototype implemented in Python, which aims to
compile ONNX models and execute them in a CM accelerator simulator that acts as
the target platform. While still work in progress, our implementation is
available as open source in \url{https://github.com/IBM/cmnnc}.

\subsection{Further Challenges}
\label{sec:challenges}

While we believe our approach shows promise, it is still at the early stages,
with missing functionality before we can claim a full-fledged solution.  Here, we
highlight some of the main challenges, which we plan to address in future work.

In our prototype, we program the LCU state machine by generating arbitrary
Python code (specifically, we generate a Python AST using the ISL AST
facilities, which we then compile to Python bytecode). This works well for our
simulation, because it enables flexibility to experiment, but a hardware
implementation might not be able to afford the ability to execute arbitrary code
on the LCU. Hence, it might be necessary to implement a more restrictive
interface for configuring the LCU (and also the GCU) so that the dependency
tracking state machine can be efficiently implemented in hardware, while
allowing the compiler to generate the appropriate configuration for arbitrary NN
dataflow graphs.

Most of the challenges of a software stack for a CM accelerator stems from the
high cost of reprogramming the crossbar arrays.  In a traditional accelerator,
if the width of a hardware unit (e.g., GEMM) cannot support the full operation,
the software can just break the operation in sub-operations and issue them
separately. In contrast, the CM accelerator has to be configured so that the
implemented dataflow execution graph deals with this limitation at
initialization time. Hence, the NN dataflow graph needs to be transformed so
that it is compatible with the CM accelerator
properties~\cite{ji2018bridge,ji2016neutrams} (e.g., performing
quantization~\cite{krishnamoorthi2018quantizing} or breaking up operations that
do not fit into individual CM cores), while, at the same time, ensuring that
the inference performance does not take a hit. As a first step to address these
challenges, we plan to quantify their effect by executing and evaluating
existing CNNs in our simulator.  \NI[We also plan to investigate if we can
train models with the constrains of the CM accelerator in mind without
significantly reducing the accuracy compared to existing models.]{I think we
should comment this out.}


\section{Related work}
\label{sec:related}

\paragraph{\bfseries Digital NN accelerators}
Most hardware NN accelerators work by considering one NN layer at a time, and
splitting its execution among multiple blocks that can be offloaded to the
device~\cite{chen2016eyeriss,chen2014diannao,nvdla,jouppi2017datacenter,chen2014dadiannao}.
Frequently, these accelerators follow a dataflow (spatial) architecture, e.g.,
by building a matrix multiply unit as a systolic
array~\cite{jouppi2017datacenter}, or by executing a 2D convolution into an
array of interconnected processing elements~\cite{chen2016eyeriss}.
Our approach, instead, fits multiple NN layers to the accelerator and
implements a dataflow execution model at a coarser granularity (inter-layer
instead of intra-layer).
One of the key challenges that recent NN accelerators try to address is how to
organize the computation to maximize data reuse within a given memory hierarchy
across the different options of reusing input data, reusing weights, and
reusing intermediate results~\cite{kwon:dnnreuse:micro19, kwon2018maeri,
sze2017efficient, fusedcnn:micro16}.  On that spectrum, the CM accelerator
described here is an extreme because the weights are directly encoded into the
crossbars.


\paragraph{\bfseries CM NN accelerators}
There is extensive work in accelerating NNs using
CM~\cite{song2017pipelayer,liu2015reno,jain2018rxnn,cheng2017time,yu2018neuro,li2017drisa,fujiki2018memory}.
Prime~\cite{chi2016prime} is a CM accelerator based on resistive memory with a
software/hardware interface similar to the one described here, as is
ISAAC~\cite{shafiee2016isaac}, where control vectors for driving state machines
are briefly mentioned but no precise description is given. These works identify
the problem of implementing logic for respecting dependencies, but provide no
solution on how the compiler can generate this logic which is the main focus of
our work. Indeed, we are not aware of any works that tackle this problem.
PUMA~\cite{ankit2019puma} implements a CM accelerator using memristor
crossbars, and defines an ISA~\cite{ambrosi2018hardware} for programming the
accelerator. In this case, the dependencies are enforced not by a state
machine, but by respecting the order of the generated instruction sequence.

\paragraph{ML compiler frameworks}
%
Initially, ML frameworks used  hand-crafted specialized routines for
implementing ML operators for each different hardware target (CPUs, GPUs,
accelerators). A number of ML compiler frameworks have since been develop that
aim to automatically generate code for these operators, but also perform
cross-operator
optimizations~\cite{rotem2018glow,leary2017xla,chen2018tvm}.
These frameworks target traditional accelerators, and focus on offloading parts
of the NN dataflow graph and exploiting data parallelism rather than pipeline
parallelism which our work focuses on.

\paragraph{Polyhedral compilation}

Polyhedral compilation enables reasoning about properties of code in the form
of nested loops with affine bounds, and have been used for many applications
including optimizations~\cite{bondhugula2008practical, kong2013polyhedral},
accelerator mapping~\cite{grosser2016polly}, program
verification~\cite{namjoshi2016loopy}, modeling
caches~\cite{bao2017analytical,gysi2019fast}, and computing bounds on IO
complexity~\cite{olivry2019automated}. Indeed, many compiler frameworks that
target NN workloads utilize polyhedral compilation
techniques~\cite{baghdadi2019tiramisu, vasilache2018tensor,
zerrell2019stripe,mlir:affine}.
These works are complementary to ours in that they deal with transforming and
scheduling loop nests so that they can be executed efficiently, while approach
targets generating control state machines for respecting data dependencies.

\section{Conclusion}
\label{sec:conclusion}

In this paper, we presented an initial approach towards a CM accelerator and
its corresponding software stack that targets efficient inference on NN
models. We focus on the problem of explicitly programming the accelerator as a
dataflow engine and discuss how the compiler can generate control state
machines that ensure that data dependencies during pipelined execution are
respected.

\newpage

\appendix
\section{Computing $\mathcal{S}$ using ISL}
\TG{Verify that all here makes sense}
\label{app:isl}

Here, we present the details of how to compute the relation $\mathcal{S}$ using
ISL.
Assuming an iteration space $I$ that writes an object $O$, and an iteration
space $J$ that reads it (\Sec{sec:deps}), we want to compute relation
$\mathcal{S} \enspace (O \rightarrow J)$ that maps observed writes in $O$, to
the maximum iteration in $J$ that can be executed.

\subsection{Background}

For completeness, we briefly present the ISL operators that we use.  We make
some simplifications for brevity, and we refer the reader to the ISL
documentation~\cite{verdoolaege2016presburger,verdoolaege2011integer} for a
complete discussion.

We represent iteration spaces ($I$, $J$) and object locations ($O$) as
{\bfseries sets} of integer tuples. We also consider {\bfseries relations}
which map elements of one set to another. For example, a relation $I
\rightarrow O$ consists of pairs ($i \rightarrow o$) such that $i \in I$ and
$o \in O$.

The domain $dom(R)$ of a relation $R$ is the set defined by the first element
of the pairs.
\[
dom(R) = { i : \exists j : (i \rightarrow j) \in R }
\]

The inverse of a relation $R$, $\mathbf{ R^{-1} }$ includes the same
tuple pairs as $R$, but with their order reversed:
\[
R^{-1} = \left\{ \enspace (j \rightarrow i) : (i \rightarrow j) \in R \enspace \right\}
\]

Relations $A$,$B$ can be composed as $\mathbf{B(A)}$ (or $B$ \emph{after} $A$):
\[
 B(A) = \left\{ i \rightarrow j : \exists k: (i \rightarrow k) \in A \ \land \ (k \rightarrow j) \in B \right\}
\]

Integer tuples can be ordered lexicographically. We use
{\boldmath$\prec$},
{\boldmath$\succ$},
{\boldmath$\preceq$},
{\boldmath$\succeq$}
to represent this order.

The lexicographical maximum $\mathbf{ \operatorname{\mathbf{lexmax}}(R)}$ of a
relation $R$ is a subset of $R$, where for pairs in $R$ with the same first element,
it only keeps the pair with the lexicographically maximal second element.

\begin{align*}
\operatorname{lexmax}{(R)} = \{ (i \rightarrow j) &: (i \rightarrow j) \in R \\
                                                        &\land \forall (i' \rightarrow k) \in R : i = i' \Rightarrow k \preceq j \} \\
\end{align*}



\subsection{Computing $S$}

Relation $W_1^{-1}$ ($O \rightarrow I$) maps each array
location to the iteration in $I$ that writes it.
Relation $\mathcal{K}$ pairs read iterations $j \in J$
write iterations $i \in I$ based on RAW dependencies.
That is, if $(j \rightarrow i) \in \mathcal{K}$, iteration $i$ writes locations
read by iteration $j$:

\[
	\mathcal{K}  :=  W_1^{-1}(R_2)  \qquad (J \rightarrow I)
\]

Next, we compute relation $\mathcal{L}$ that pairs every read iteration
$j \in J$ to the last write iteration $i \in I$ that satisfies all dependencies
for all iterations up to and including $j$. In other words, after $i$ is
executed, the reader can safely execute all iterations until $j$. Note that the
above assumes an order between iterations. We assume iterations are executed in
their lexicographical order, as is normally the case.
We compute $\mathcal{L}$ as follows. First, we compute the domain
$D$ of $\mathcal{K}$.

\[
  D  := \operatorname{dom}{(\mathcal{K})}  \qquad (J)
\]

Next, we use the \emph{lexicographically-greater-than-or-equal} relation on
sets operations ($>>=$) to compute relation $D'$ such that each iteration $j$
is mapped to all iterations $\zeta$ that do not succeed it ($\zeta \preceq j$).

\[
	D' :=   D >>= D   \qquad (J \rightarrow J)
\]

We can apply relation $D'$ after $\mathcal{K}$, which results in a
relation that pairs each read iteration $j$ to every write iteration $i$ that
writes data read by iterations $\zeta \preceq j$. We compute $\mathcal{L}$
using the \emph{lexmax} operator, which will only keep the last write iteration
after which all iterations $\zeta \preceq j$ can be safely executed.

\[
	\mathcal{L}  :=  \operatorname{lexmax}{ \left( \mathcal{K}(D') \right) }  \qquad (J \rightarrow I) \\
\]

Applying $W_1$ \emph{after} $\mathcal{K}$, results in relation $\mathcal{M}$,
which if reversed maps locations written by the writer loop, to the maximum
iteration we can execute on the reader loop. Because a single write might be
mapped to multiple read iterations, we use the \emph{lexmax} operator to keep
the maximal (i.e., latest).

\begin{alignat*}{2}
	\mathcal{M}  & :=   W_1(\mathcal{L}) &  (J \rightarrow O) \\
	\mathcal{S}  & :=   \operatorname{lexmax}{ \left( \mathcal{M}^{-1} \right) }  \qquad &  (O \rightarrow J) \\
\end{alignat*}

\newpage

\bibliography{references}

\end{document}